\documentclass[11pt]{iopart}
\usepackage{bm}
\usepackage{hyperref}
\usepackage{amssymb}
\usepackage{amsopn}

\usepackage{subfigure}
\usepackage{graphicx}

\newcommand{\be}{\begin{equation}}
  \newcommand{\ee}{\end{equation}}
\newcommand{\ba}{\begin{eqnarray}}
  \newcommand{\ea}{\end{eqnarray}}

\newcommand{\bmp}{\bar{m}_p}

\newcommand{\de}{\delta}

\newcommand{\Om}{\Omega}
\newcommand{\lam}{\lambda}
\newcommand{\omd}{\Omega_{d c}}
\newcommand{\omc}{\Omega_{c c}}
\newcommand{\oco}{\Omega_{c 0}}
\newcommand{\obo}{\Omega_{b 0}}
\newcommand{\odo}{\Omega_{d 0}}
\newcommand{\oro}{\Omega_{r 0}}
\newcommand{\zcmb}{z_{\rm CMB}}

\renewcommand{\(}{\left(}
\renewcommand{\)}{\right)}
\renewcommand{\[}{\left[} 
\renewcommand{\]}{\right]}

\begin{document}

\title{The coincidence problem and interacting holographic dark energy}
\author{Khamphee Karwan}
\address{Theoretical High-Energy Physics and Cosmology Group,
Department of Physics, Chulalongkorn University, Bangkok 10330, Thailand
}

\begin{abstract}
We study the dynamical behaviour of the interacting holographic dark energy model
whose interaction term is $Q=3H(\lam_d\rho_d + \lam_c\rho_c)$,
where $\rho_d$ and $\rho_c$ are the energy density of dark energy and CDM respectively.
To satisfy the observational constraints from SNIa, CMB shift parameter and BAO measurement, if $\lam_c = \lam_d$ or $\lam_d, \lam_c >0$,
the cosmic evolution will only reach the attractor in the future and the ratio $\rho_c/\rho_d$ cannot be slowly varying at present.
Since the cosmic attractor can be reached in the future even when the present values of the cosmological parameters do not satisfy the observational 
constraints, the coincidence problem is not really alleviated in this case.
However, if $\lam_c \neq \lam_d$ and they are allowed to be negative,
the ratio $\rho_c/\rho_d$ can be slowly varying at present and the cosmic attractor can be reached near the present epoch.
Hence, the alleviation of the coincidence problem is attainable in this case.
The alleviation of coincidence problem in this case is still attainable when confronting this model to SDSS data.

\vspace{3mm}
\begin{flushleft}
\end{flushleft}
\end{abstract}

\maketitle

\section{Introduction}

Observations suggest that the expansion of the universe is accelerating \cite{Riess:98,Perl:99}.
The acceleration of the universe may be explained by supposing that 
the present universe is dominated by a mysterious form of energy whose pressure is negative, known as dark energy.
One problem of the dark energy model is the coincidence problem,
which is the problem why the dark energy density and matter density are of the same order of magnitude in the present epoch although they differently evolve during the expansion of the universe.
A possible way to alleviate the coincidence problem is to suppose that there is an interaction between matter and dark energy.
The cosmic coincidence can then be alleviated by appropriate choice of the form of the interaction between matter and dark energy
leading to a nearly constant ratio $r=\rho_c/\rho_d$ during the present epoch
\cite{Zimdahl:03, Campo:06, Sadjadi:06}
or giving rise to attractor of the cosmic evolution at late time \cite{Amen:99, Chimento:03}.
Since the existence of the cosmic attractor implies constant $r$ but the attractor does not always occur at the present epoch,
we first find a range of dark energy parameters for which the attractor occurs and then check the evolution of $r$ during the present epoch.

Based on holographic ideas \cite{Cohen:98, Li:04}, one can determine the
dark energy density in terms of the horizon radius of the universe. This type of dark energy is holographic dark energy
\cite{Setare:07} - \cite{Elizalde:05}.
By choosing Hubble radius as the cosmological horizon, the present amount of dark energy density agrees with observations.
Nevertheless, dark energy evolves like matter at present, so
it cannot lead to an accelerated expansion.
However, if the particle horizon is chosen to be the cosmological horizon,
the equation of state parameter of dark energy can become negative
but not negative enough to drive an accelerating universe.
The situation is better when one uses the event horizon as the cosmological horizon.
In this case, dark energy can drive the present accelerated expansion,
and the coincidence problem can be resolved by assuming an appropriate number of e-foldings of inflation.
Roughly speaking, the coincidence problem can be resolved because the size of the cosmological horizon during the present epoch depends on
the amount of e-folds of inflation, and the amount of holographic dark energy depends on the horizon size.
Nevertheless, the second law of thermodynamics will be violated if $w_d < -1$ \cite{Li:04, Gong:06}.
Hence, $w_d$ should not cross the boundary $w_d = -1$.
The boundary $w_d = -1$ can be crossed if dark energy interacts with matter. Since now the horizon size has a dependence on the interaction 
terms,the alleviation of cosmic coincidence should also depend on the interaction term.

In this work we suppose that the holographic dark energy interact only with cold dark matter (CDM) and treat baryons as non-interacting matter 
component. Our objective is to compare the region of dark energy parameters for which the cosmic evolution has an attractor within the parameter region 
that satifies the observational constraints from combined analysis of SNIa 
data \cite{Riess:06}, CMB shift parameter \cite{Wang:06} and BAO measurement
 \cite{Eisenstein:05}. The results of the comparison can tell us about the range of parameters that alleviate the cosmic coincidence.

\section{The autonomous equations}

In this section, we derive the first order differential equations that describe the evolution of radiation, baryon, CDM 
and dark energy densities in the universe.
By analyzing these equations, one can estimate the asymptotic evolution of the universe.
To proceed, we start from the Friedmann equation
\be
H^2 + \frac{K}{a^2} = \frac 1{3\bmp} \(\rho_r + \rho_b + \rho_c + \rho_d\),
\label{h}
\ee
where $H$ is the Hubble parameter and
the subscripts $r, b, c$ and $d$ correspond to the radiation, baryons, CDM and dark energy respectively.
The parameter $K$ denotes the curvature of the universe,
where $K=-1, 0 , +1$ for the close, flat and open universe respectively.
The above equation can be written in terms of the density parameters
$\Om_K=K/(a^2H^2)$ and  $\Om_\alpha = \rho_\alpha/(3\bmp^2H^2)$ as
\be
1 + \Om_K = \sum_{\alpha = r, b, c, d}\Om_\alpha
=  \Om_r + \Om_b + \Om_c + \Om_d.
\label{h2}
\ee
The index $\alpha$ runs over the 4 species, namely radiation, baryons, CDM and dark energy.
We now derive the autonomous equations for the dynamical variables
$\Om_K$ and $\Om_\alpha$.
Differentiating $\Om_K=K/(a^2H^2)$ with respect to $\ln a$, we get
\be
\Om_K' = -\frac{2K}H \(\frac{\dot a}{a^3H^2} + \frac{\dot H}{a^2H^3}\)
= -2\Omega_K\(1+\frac{\dot H}{H^2}\),
\label{omk}
\ee 
where prime and dot denote  derivative with respect to $\ln a$ and time respectively.
From the definition of the density parameter, one can show that 
\be
\Om_\alpha' = \Om_\alpha\(\frac{\dot\rho_\alpha}{H\rho_\alpha} - 2\frac{\dot H}{H^2}\).
\label{oma}
\ee
To study the evolution of the universe at late time,
we will search for the fixed points of the above autonomous equations
and check the stability of these fixed points.
The fixed points of eqs. (\ref{omk}) and (\ref{oma}) are the points 
$(\Om_{K c}, \Om_{\alpha c})$ at which
\be
\Om_K' = \Om_\alpha' = 0.
\label{dffix}
\ee
It follows from eq. (\ref{omk}) that $\Om_K' = 0$
at $\Om_K=0$ or $1+\dot H/H^2 = 0$.
Hence, possible fixed points at $\Om_K\neq 0$ correspond to the non-accelerating universe,
i.e. $1+\dot H/H^2 \propto \ddot a = 0$.
Since the expansion of the universe is accelerating today,
we consider only the fixed points at $\Om_K = 0$,
and therefore neglect $\Om_K$ in our consideration for simplicity.

Now we come to the case of interacting holographic dark energy and use eq. (\ref{oma}) to obtain the autonomous equation for this case.
In the holographic dark energy scenario, the energy density of dark energy is related to the cosmological horizon $L$ by
\be
\rho_d = 3c^2\bmp^2L^{-2},
\label{holo}
\ee
where $c$ is a positive constant.
Differentiating the above equation with respect to time, we obtain
\be
\dot\rho_d = -2\rho_d\frac{\dot L}{L}.
\label{drd1}
\ee
We take the cosmological horizon to be the event horizon,
which is defined as $R_e(t) = a(t)\int_t^\infty d\tilde{t}/a(\tilde{t})$.
Hence, 
\be
\frac{\dot L}{L} = \frac{\dot R_e}{R_e} = H - \frac 1{R_e}.
\ee
We therefore get
\be
\dot\rho_d = -2H\rho_d + 2\frac{\rho_d^{3/2}}{\sqrt{3} c \bmp}.
\label{drdho}
\ee
When dark energy has an interaction with CDM, the continuity equations yield
\ba
\dot\rho_c &=& -3H\rho_c +Q,
\label{drm2}\\
\dot\rho_d &=& -3H(1+w_d)\rho_d - Q,
\label{drd2}
\ea
where $Q = 3H(\lam_d\rho_d + \lam_c\rho_c)$.
Usually, one supposes that $Q>0$ because the second law of thermodynamics might be violated
if energy transfers from matter to dark energy ($Q<0$).
However, for generality, we will not restrict $Q$ to be positive in our consideration. 
Comparing eq. (\ref{drdho}) with eq. (\ref{drd2}), we get
\be
w_d = -\frac 13 - 2\frac{\sqrt\Om_d}{3c} - \frac{\lam_d\Om_d + 
\lam_c\Om_c}{\Om_d}.
\label{wd}
\ee
We assume that radiation and baryons have no interaction with dark energy, so that
they obey the continuity equations
\be
\dot\rho_r = -4H\rho_r \quad\quad\mbox{and}\quad\quad \dot\rho_b = -3H\rho_b.
\label{dr}
\ee
From eq. (\ref{h}), one can show that
\be
2\frac{\dot H}{H^2} = \frac{\(H^2\)^{.}}{H^3}
= -3 + \Om_d + 2\frac{\Om_d^{3/2}}{c} + 3(\lam_d\Om_d + \lam_c\Om_c) - \Om_r.
\label{dh}
\ee
In the above equation, we set $\Om_K =0$.
Using eqs. (\ref{oma}), (\ref{drdho}), (\ref{dr}) and (\ref{dh}), we obtain
\ba
\Om_d' &=& \Om_d\(1+2\frac{\Om_d^{1/2}}{c} - \Om_d - 2\frac{\Om_d^{3/2}}{c} - 3(\lam_d\Om_d+\lam_c\Om_c) + \Om_r\),
\label{omd} \\
\Om_c' &=& \Om_c\(\frac{3}{\Om_c}(\lam_d\Om_d + \lam_c\Om_c) - \Om_d - 2\frac{\Om_d^{3/2}}{c} - 3(\lam_d\Om_d+\lam_c\Om_c) + \Om_r\),
\label{omm}\\
\Om_r' &=& \Om_r\(-1 - \Om_d - 2\frac{\Om_d^{3/2}}{c} - 3(\lam_d\Om_d+\lam_c\Om_c) + \Om_r\),
\label{omr}\\
\Om_b' &=& \Om_b\(- \Om_d - 2\frac{\Om_d^{3/2}}{c} - 3(\lam_d\Om_d+\lam_c\Om_c) + \Om_r\).
\label{omb}
\ea
The fixed points of the above equations are $(\Om_d, \Om_c, \Om_r, \Om_b) = (0, 0, \Om_{r c}, 0)$,
$(0, 0, 0, \Om_{b c})$ and $(\omd, \omc, 0, 0)$.
Since we are interested in the late time evolution of the universe,
we will consider only the fixed point $(\omd, \omc, 0, 0)$.
This fixed point can occur at late time, i.e., about the present or in the future, because
$\Om_r/\Om_c$ usually decreases with time and $\Om_b/\Om_c$ can decrease with time if $Q>0$.
 Using eqs. (\ref{omd}) and (\ref{omm}), the relation between $\Om_{d c}$ and $\Om_{c c}$ can be written as
\be
1 + 2\frac{\omd^{1/2}}{c} = \frac{3}{\omc}(\lam_d\omd + \lam_c\omc).
\label{fix}
\ee
From eq. (\ref{h2}), we have $\omd + \omc = 1$ and hence
\be
1 - 3\lam_c + 2\frac{\omd^{1/2}}{c} - (1 + 3\lam_d - 3\lam_c)\omd
- 2\frac{\omd^{3/2}}{c} = 0.
\label{od3}
\ee
The solution of eq. (\ref{od3}) gives $\omd$ in terms of $\lam_d, \lam_c$ and $c$.
Instead of finding the solution of this equation,
we will use  eq. (\ref{fix}) to compute the cosmological parameters of interest in the following section.
The basic idea is that there are various values of
$c, \lam_d$ and $\lam_c$ that satisfy eq. (\ref{fix}) for  given $\omd$ and $\omc$.
 Changes in the values of $c, \lam_d$ and $\lam_c$ lead to a different cosmological evolutions,
i.e. a different $w_d$ for a given $\omd$ and $\omc$.

The stability of this fixed point can be investigated by linearizing eqs. (\ref{omd}) - (\ref{omb})
around the fixed point and studying how the fluctuations around the fixed point evolve in time.
If the amplitude of the fluctuations decreases in time,
the fixed point is a stable fixed point or attractor.
Linearizing eqs. (\ref{omd}) - (\ref{omb}) around $(\omd, \omc, 0, 0)$, we get
\ba
&&\de\Om_d' = \(\frac{\omd^{1/2}}{c} - \omd - 3\frac{\omd^{3/2}}{c} - 3\lam_d\omd\)\de\Om_d
- 3\lam_c\omd\de\Om_c + \omd\de\Om_r,
\label{lod}\\
&&\de\Om_c' = \(-1 - 2\frac{\omd^{1/2}}{c} + 3\lam_c\omd\)\de\Om_c
+ \(3\lam_d - \omc - 3\omc\frac{\omd^{1/2}}{c} - 3\lam_d\omc\)\de\Om_d + \omc\de\Om_r,
\nonumber\\
\label{lom}\\
&&\de\Om_r' = -2\(1+\frac{\omd^{1/2}}{c}\)\de\Om_r,
\label{lor}\\
&&\de\Om_b' = -\(1+2\frac{\omd^{1/2}}{c}\)\de\Om_b,
\label{lob}
\ea
where $\de\Om_\alpha = \Om_\alpha - \Om_{\alpha c}$ denote  fluctuations around the fixed point.
The above linear equations can be written as
\be
\de\Om_{\alpha}' = M_{\alpha\beta}\de\Om_\beta,
\ee
where $\alpha$ and $\beta$ run over the 4 species.
The eigenvalues of the matrix $M$ govern how the amplitude of the fluctuations around the fixed point changes with time.
The fixed point is a stable, saddle or unstable point if all the eigenvalues are negative, some of eigenvalues are positive or all the eigenvalues are positive,respectively.(Since we are dealing with real eigenvalues.)
The eigenvalues of the matrix $M$ are
\ba
\lam_1 &=& \frac{\omd^{1/2}}{c} - (1 + 3\lam_d - 3\lam_c)\omd - 3\frac{\omd^{3/2}}{c}, \nonumber\\
\lam_2 &=& \lam_4 = -1-2\frac{\omd^{1/2}}{c}, \quad\quad
\lam_3 = -2\(1+\frac{\omd^{1/2}}{c}\).
\ea
Hence, the fixed point $(\omd, \omc, 0, 0)$ is a stable point when $\lam_1<0$ and a saddle point when $\lam_1>0$. 
Using eq. (\ref{fix}), $\lam_1$ can be written as
\be
\lam_1 = \frac{3(\lam_d\omd+\lam_c\omc)}{\omc}\(\frac 12 - \omd\) - \frac 12 - \frac{\omd^{3/2}}{c} - 3(\lam_d-\lam_c)\omd.
\label{lam}
\ee
It is not easy to determine the sign of $\lam_1$ in general.
Thus, we will determine it in particular cases in the next section.

\section{The attractor of cosmic evolution}

We now consider the fixed point and stability of the cosmic evolution around the present epoch.
For simplicity, we first consider the case where $\lam_d = \lam_c = b^2$.
In this case, eq. (\ref{fix}) becomes
\be
1+2\frac{\omd^{1/2}}{c} = \frac{3b^2}{\omc}.
\label{e1}
\ee
Since $c>0$, eq. (\ref{e1}) implies that $b^2\geq\omc / 3$.
This is the lower limit of $b^2$ for the existence of the fixed point.
From eqs. (\ref{wd}) and (\ref{e1}), it is easy to show that
\be
w_d = -\frac{b^2}{\omc\omd}.
\label{e1wd}
\ee
The equation of state parameter of the universe is defined as
\be
 w = \frac{p_{\rm total}}{\rho_{\rm total}} = \sum_{\alpha = r, b, m, d}w_\alpha\Om_\alpha,
\ee
where $w_\alpha = p_\alpha / \rho_\alpha$.
Since $w_b=w_c=0$ and $\Om_r$ can be neglected at late time,
we have $w = w_d\Om_d$ and therefore.
\be
w = -\frac{b^2}{\omc}.
\ee
From the lower bound of $b^2$, we get $w \leq 1/3$.
This means that the fixed point corresponds to  cosmic acceleration.
Moreover, one can see that the universe will be in the phantom phase, i.e., $w<-1$,
if $b^2>\omc$ or equivalently $\omd^{1/2} > c$.
Since the observations seem to indicate that $w_d<-1$ today,
we find the value of $b^2$ that makes $w_d<-1$ at the fixed point.
It follows from eq. (\ref{e1wd}) that $w_d<-1$ if $b^2>\omc\omd$.
We now check the stability of the fixed point.
In this case, eq. (\ref{lam}) becomes
\be
\lam_1 = \frac{3b^2}{\omc}\(\frac 12 - \omd\) - \frac 12 - \frac{\omd^{3/2}}{c}.
\ee
It can be seen that it is not easy to find a point at which $\lam_1$ changes sign.
However, $\lam_1$ is negative for any $b^2$ or $c$ if $\omd > 1/2$,i.e. $\omd > \omc$.
This means that the fixed point at which $\omd > \omc$ is a stable fixed point or attractor.
Since the fixed point will occur when $Q$ is positive and $\Om_r = \Om_b =0$. At the present epoch  $\Om_r$ is small and 
can be neglected  and with positive $Q$ the ratio $\Om_b / \Om_c$ decreases with time, so
 the fixed point $(\omd, \omc, 0, 0)$ will be reached in the future.
From eqs. (\ref{omd}) - (\ref{omb}), one can see that
if $b^2$ is  large,
$\Om_b$ can decrease quickly with time compared with $\Om_c$.
As a result, the fixed point can be reached quickly near the present epoch.
However, the CDM dominated epoch will disappear and the baryon fraction will be larger than the CDM fraction
in the last scattering epoch due to the rapid decrease of $\rho_b/\rho_c$ with time.
This is excluded by the observed peak height ratio of the CMB power spectrum.
Thus, based on observations, the fixed point cannot be reached near the present for this case.
We will see in the next section that small $b^2$ or equivalently small $\omc$ is required by observations.
Substituting $\omc = 1 - \omd$ into eq. (\ref{e1}), we get
\be
1 - 3b^2 + 2\frac{\omd^{1/2}}{c} - \omd - 2\frac{\omd^{3/2}}{c} = 0
\ee
The above third degree polynomial for $\omd^{1/2}$ will have one positive real root
if $1 > 3b^2$. This root is the previously considered fixed point.
In contrast, if $1 \leq 3b^2$ the above equation will have one additional real root.
This root gives another fixed point at smaller $\omd$.
It is not hard to show that this fixed point is not  stable and we will not consider it here.

Let us now consider the case where $\lam_d \neq \lam_c$.
In this case, eq. (\ref{fix}) becomes
\be
1+2\frac{\omd^{1/2}}{c} = \frac{3}{\omc}\(\lam_d\omd + \lam_c\omc\).
\label{e3}
\ee
Hence, the fixed point occurs when $\lam_d\omd + \lam_c\omc > \omc / 3$.
Using eqs. (\ref{wd}) and (\ref{e3}), we get
\be
w_d = -\frac{\lam_d\omd + \lam_c\omc}{\omc\omd},
\label{e3wd}
\ee
and therefore
\be
w = -\frac{\lam_d\omd + \lam_c\omc}{\omc}.
\label{e3w}
\ee
Similarly to the case when $\lam_d = \lam_c$, one can show that the fixed point occurs
when $w_d < -1/(3\omd )$ or $w < - 1/3$.
Since the terms $\lam_d\omd$ and $\lam_c\omc$ from the interaction dominate at different times,
the parameters $\lam_d$ and $\lam_c$ can be chosen such that the baryon fraction is smaller than the CDM fraction during the last scattering epoch
although $\omc$ need not be very small.
Thus, in this case, the attractor can be reached faster than the case of $\lam_c = \lam_d$.
From eq. (\ref{lam}), we see that the fixed point is the stable fixed point when $\omd > \omc$ and $\lam_c$ is not much larger than $\lam_d$.

\section{The observational constraints}

We now check the range of parameters for which the cosmic evolution has an attractor for compatibility with  observations.
We constrain the parameters of the interacting holographic dark energy using the latest observational data of SNIa \cite{Riess:06} combined
with the CMB shift parameter derived from three-year WMAP \cite{Wang:06} and the baryon acoustic
oscillations (BAO) from SDSS LRG \cite{Eisenstein:05}.

The SNIa observations measure the apparent magnitude $m$ of
a supernova and its redshift $z$. The apparent magnitude $m$ is
related to the distance modulus $\mu$ and luminosity distance $d_L$ of the supernova by
\be
\mu(z) = m(z) - M = 5\log_{10}(d_L(z)/{\rm Mpc})+25,
\ee
where $M$ is the absolute magnitude of the supernova.
For flat space time, the luminosity distance is given by
\be
d_L(z) = H_0^{-1}(1+z)\int_0^z{d\tilde{z}\over E(\tilde{z})}.
\ee
Here, $H_0$ is the present value of the Hubble parameter and
$E(z) = H(z)/H_0$. To constrain the holographic dark energy model, we perform $\chi^2$ fit for
the parameters $\lam_d, \lam_c, c, \oco, \obo$, where the subscript $0$ denotes the present value.
 For $\oro$, we compute its value from the CMB and Neutrino temperatures.
Using eq. (\ref{h2}), one can compute $\odo$ from $\oco, \obo$ and $\oro$. We have to include radiation in our consideration
because it should not be neglected when we compute CMB shift parameter. For the SNIa data, the parameter $H_0$ is the 
nuisance parameter which needs to be marginalized out.
Since the fit of holographic dark energy to SNIa data is sensitive to $H_0$ \cite{Zhang:05},
we have to add constraints from other observations to improve the fit.
For this reason, we include the constraints from CMB shift parameter and BAO measurement in our analysis.

The CMB shift parameter is a quantity derived from CMB data that has been 
 shown to be model independent, so it can be used to constrain cosmological models.
The CMB shift parameter is defined as \cite{Zhang:07}
\be
R = \Omega_{m0}^{1/2}\int_0^{\zcmb}{d\tilde{z}\over E(\tilde{z})},
\ee
where $\Om_{m 0} = \oco + \obo$ and $\zcmb = 1089$ is the redshift at recombination.
The estimated value of $R$ from 3-year WMAP data is
$1.70\pm 0.03$.
From the SDSS data, the measurement of the BAO peak in the distribution of SDSS LRG
can be used to derive the model independent parameter $A$ which is defined as \cite{Eisenstein:05}
\be
A = \Omega_{m0}^{1/2} E(z_{\rm BAO})^{-1/3}\[{1\over z_{\rm BAO}}\int_0^{z_{\rm BAO}}{d\tilde{z}\over E(\tilde{z})}\]^{2/3},
\ee
where $z_{\rm BAO}=0.35$.
The estimated $A$ is
$A=0.469(n_s/0.98)^{-0.35}\pm 0.017$.
According to the 3-year WMAP data, the scalar spectral index is chosen to be $n_s=0.95$.

We first consider the case where $\lam_d=\lam_c=b^2$.
For simplicity, we start by neglecting baryons in our consideration.
Hence the attractor becomes $(\Om_d, \Om_c, \Om_r) = (\omd, \omc, 0)$, where $\omd$ and $\omc$ satisfy eq. (\ref{e1}).
The region of $b^2$ and $\omc$ for which the fixed point exists and
the $99.7\%$ confidence regions from the combined constraints are shown in figure 1.
From this figure, one sees that a small $b^2$ is required by observations,
so that the attractor that satisfies the observational constraints occurs at low $\omc$.
This implies that the attractor cannot occurs at present at which $\Om_c \approx 0.3$.
In the future, $\Om_c$ can become small, so that the fixed point can exist.
Nevertheless, if the attractor is reached in the far future, the cosmic coincidence may not be alleviated
because the present value of $\Om_c$ and $w_d$ may not satisfy the observational constraints.
In the following consideration, we will study how the ratio $r$ evolves during the present epoch
to check the possibility of alleviating the cosmic coincidence.
Since the physical fixed point cannot occur if $b^2 < 0$, we perform another fit by supposing that $b^2 \geq 0$.
The result is shown in figure 1. It can be seen that the above conclusions for the case of arbitrary $b^2$ are also 
valid for the case of $b^2\geq 0$.
\begin{figure}[ht]
\includegraphics[height=0.4\textwidth, width=0.8\textwidth,angle=0]{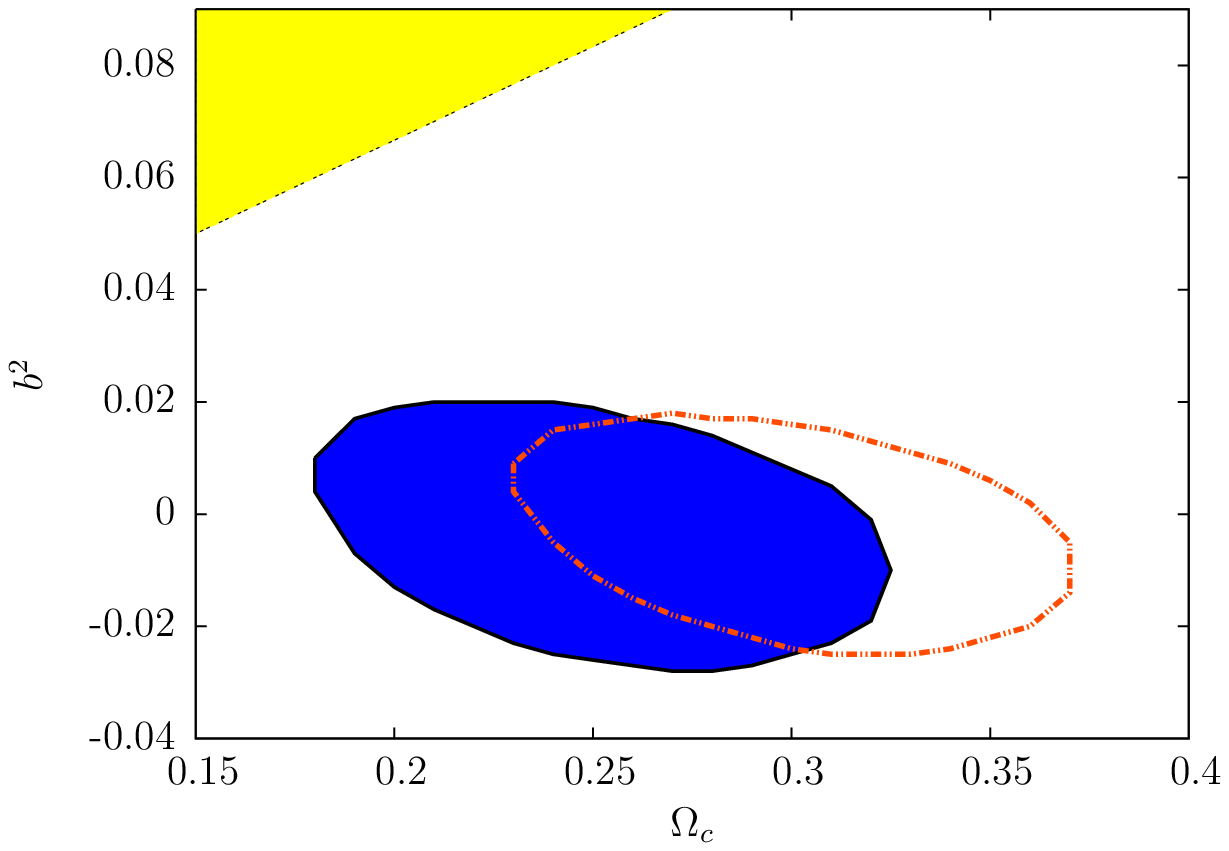}
\\
\includegraphics[height=0.4\textwidth, width=0.8\textwidth,angle=0]{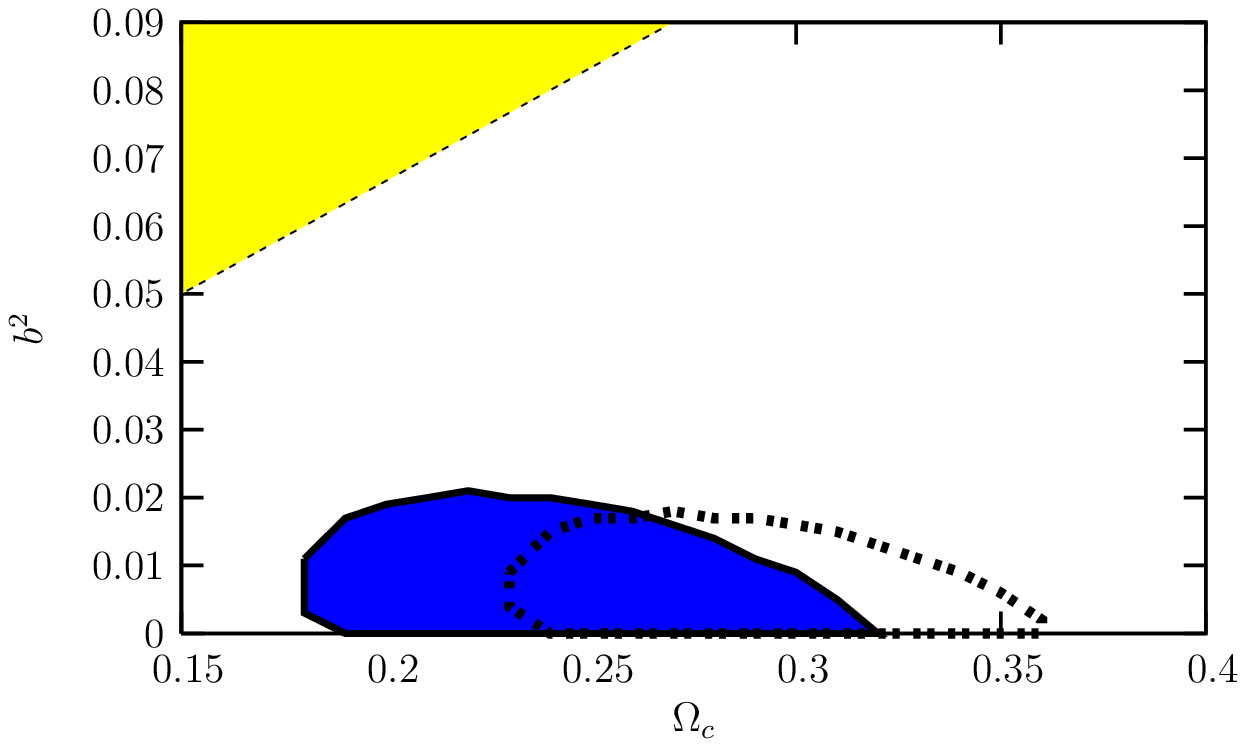}
\caption{
A region of parameters $b^2$ and $\Om_c$ in which the cosmic evolution has a late time attractor
(the yellow regions above the dashed lines),
and the $99.7\%$ confidence levels from the combined analysis of SNIa data , CMB shift parameter and BAO measurement.
For the yellow region, $\Om_c$ refers to $\omc$, but
$\Om_c$ refers to $\oco$ for the confidence levels.
The upper panel shows the case of arbitrary $b^2$, while
the lower panel shows the case where $b^2\geq 0$.
The blue regions represent the confidence regions for the case that includes baryons and prior $\obo = 0.047\pm 0.006$,
while the thick dotted lines represent the confidence levels for the case where baryons are neglected.
}
\label{fig::1}
\end{figure}

The situation changes a bit when we include baryons in our consideration.
From the previous section, we know that the cosmic evolution reaches the attractor at $(\Om_d, \Om_c, \Om_b, \Om_r) = (\omd, \omc, 0, 0)$.
Since the present value of $\Om_b$ does not vanish, this attractor cannot occur at present.
However, the attractor can occur in the future because the ratio $\Om_b / \Om_c$ decreases with time due to the positive $Q$.
It can be seen that the ratio $\Om_b/ \Om_c$ decreases faster when $b^2$ increases.
According to the observational constraints, a small $b^2$ is also required in this case,
so that the attractor is slowly reached in the future.
In order to perform a fit for this case, we use a prior $\obo = 0.047 \pm 0.006$ from the 1-year WMAP data \cite{wmap1:03}
because $\obo$ cannot be constrained very well by using only SNIa data, CMB shift parameter and BAO measurement.
It can be seen from figure 1 that the shape of the confidence contours does not change much when we include non-interacting baryons in our consideration.
Obviously, the contours move to the left when the amount of $\obo$ increases.
This is because the amount of $\oco$ decreases.
\begin{figure}[ht]
 \includegraphics[height=0.4\textwidth, width=0.8\textwidth,angle=0]{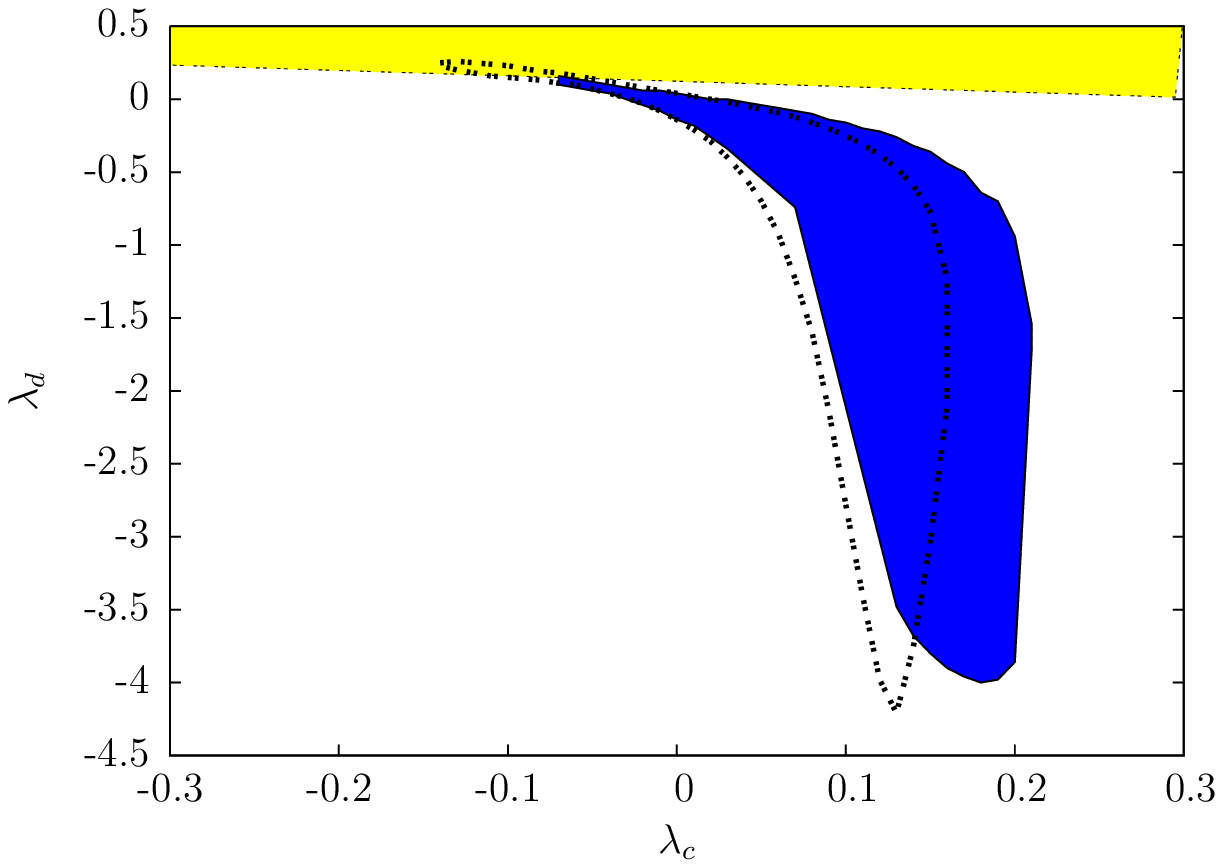}
\\
\includegraphics[height=0.4\textwidth, width=0.8\textwidth,angle=0]{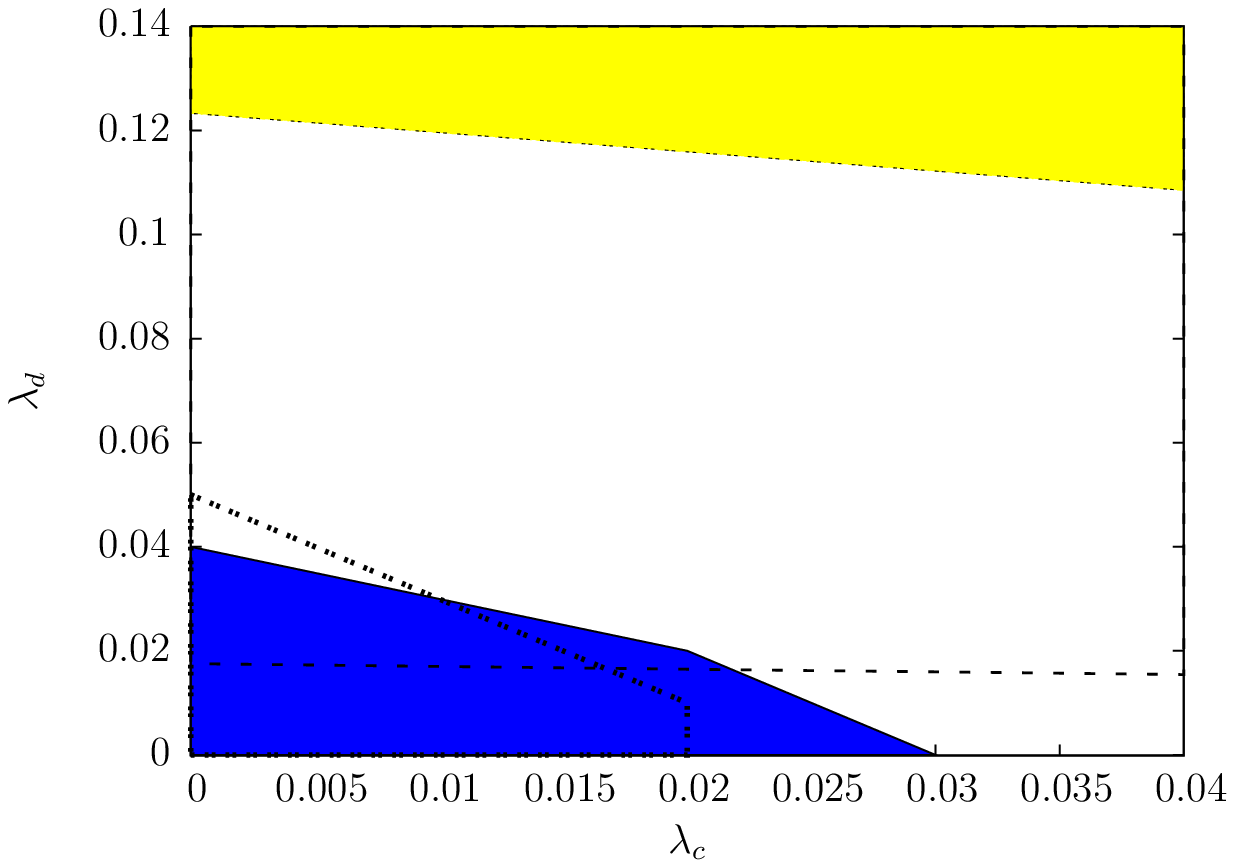}
\caption{
A region of parameters $\lam_d$ and $\lam_c$ for which the cosmic evolution has a late time attractor
(the yellow regions above the dashed lines),
and the $99.7\%$ confidence levels from the combined analysis of SNIa data , CMB shift parameter and BAO measurement.
The upper panel shows the case of arbitrary $\lam_d$ and $\lam_c$, while
the lower panel shows the case where $\lam_d, \lam_c\geq 0$.
The blue regions represent the confidence regions for the case that includes baryons and prior $\obo = 0.047\pm 0.006$,
while the thick dotted lines represent the confidence levels for the case where baryons are neglected.
The dashed lines are plotted by setting $\Om_{c c} = 0.27$,
but the thick dashed line in the lower panel is plotted by setting $\Om_{c c} = 0.05$.
We note that the region above the thick dashed line also represents the region for which the cosmic evolution has attractor.
}
\label{fig::2}
\end{figure}

We now consider the case of arbitrary $\lam_d$ and $\lam_c$.
We also perform a $\chi^2$ fit for the case with baryons and without baryons.
For the case where baryons are neglected, it follows from figure 2 that the attractor can occur at
present ($\Om_{c c} \approx 0.3$) for a narrow range of $\lam_d$ and $\lam_c$.
Nevertheless, a small $\Om_{c c}$ is required by observations if we restrict $\lam_d$ and $\lam_c$ to be positive.
This implies that the attractor must occur in the future.
Similarly to the case of $\lam_d = \lam_c = b^2$, the attractor cannot be reached at present
if baryons are included in the consideration.
Nevertheless, for suitable values of $\lam_c$ and $\lam_d$ which satisfy the observational constraints, the attractor in this case can be reached faster than the case of $\lam_c = \lam_d$.
\begin{figure}[ht]
\includegraphics[height=0.4\textwidth, width=0.8\textwidth,angle=0]{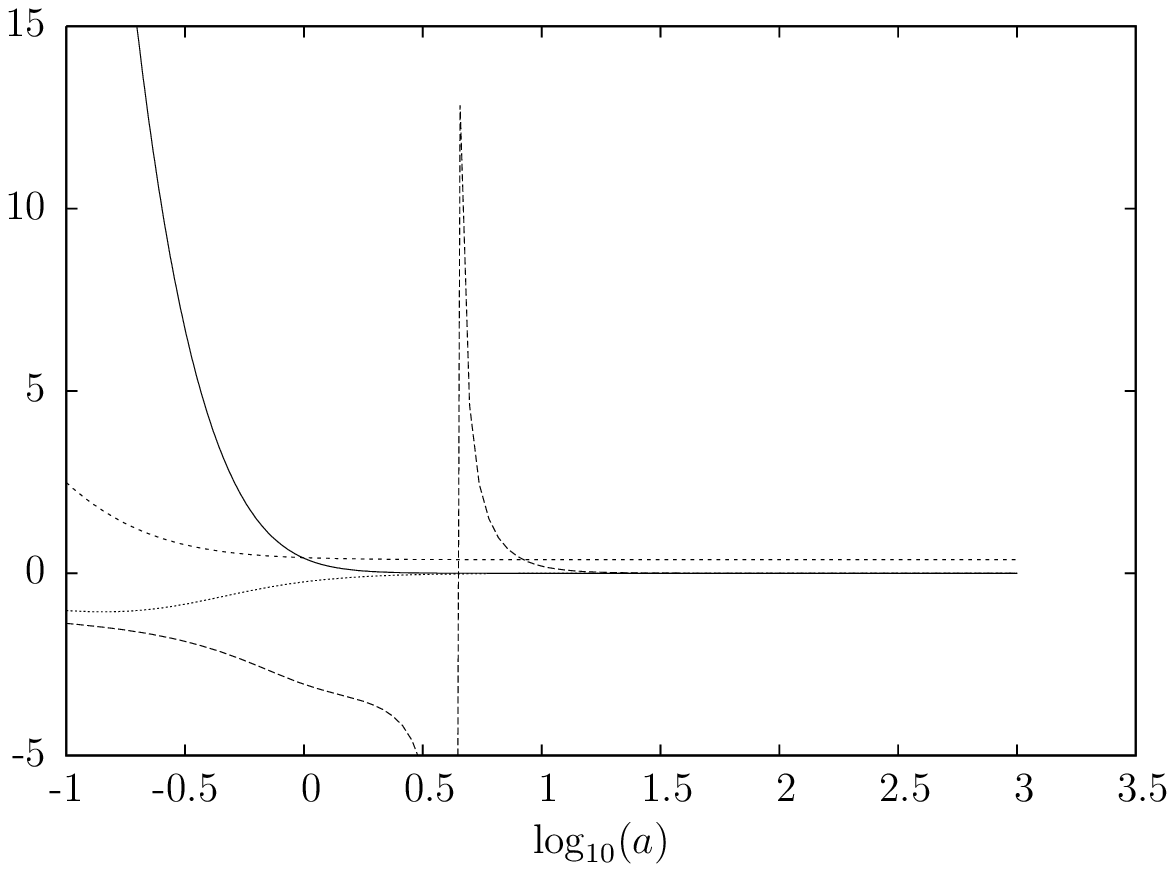}
\includegraphics[height=0.4\textwidth, width=0.8\textwidth,angle=0]{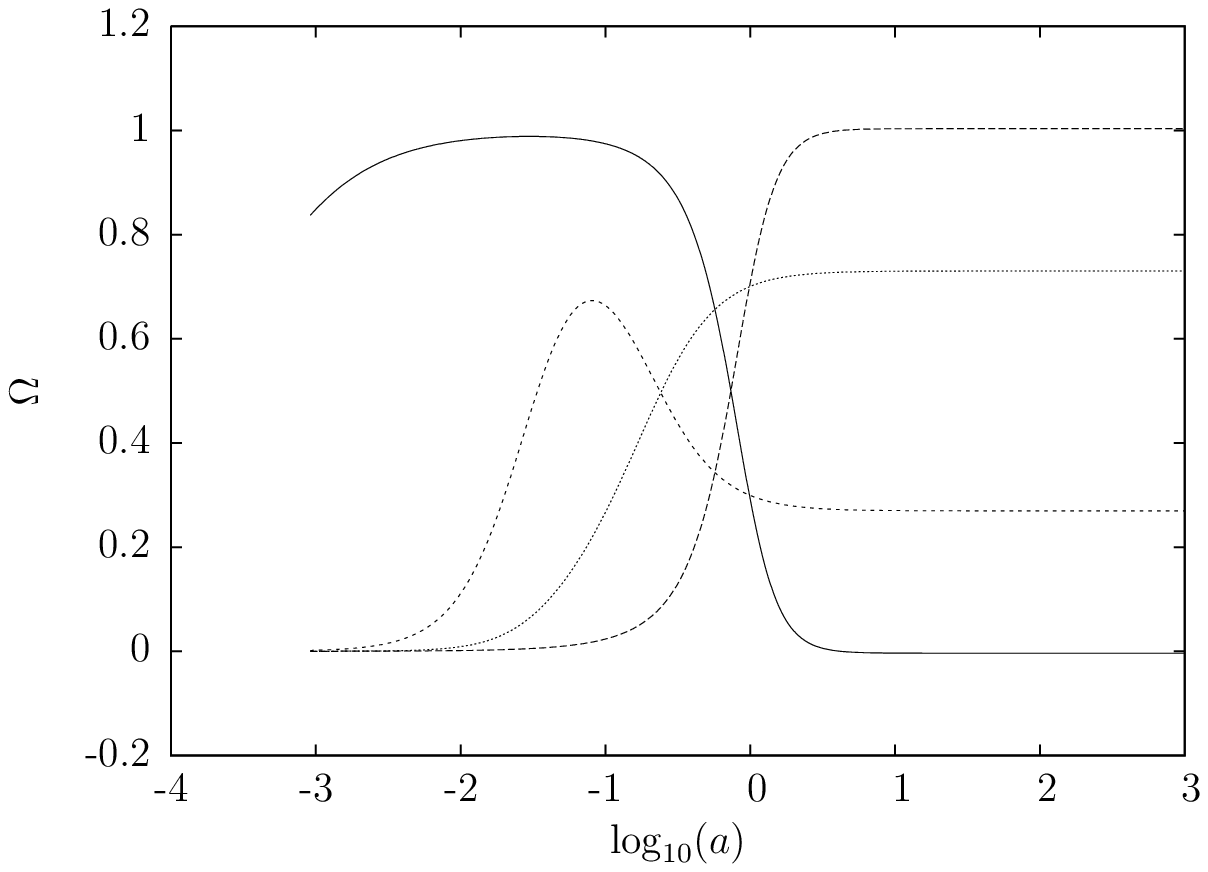}
\caption{
The upper panel shows the evolution of $r$ (solid and dashed lines) and $\dot{r}/(r H)$ (long dashed and dotted lines),
while the lower panel shows the evolution of $\Om_c$ (solid and dashed lines) and $\Om_d$ (long dashed and dotted lines).
The solid and long dashed lines correspond to the case where the value of $b^2$, $c$ and $\Om_{c 0}$ equal to their best fit value,
while the dashed and dotted lines correspond to the case where $b^2$ is chosen such that $\Om_{c c}$ is close to $\Om_{c 0}$,i.e. $\Om_{c c} = 0.27$.
}
\label{fig::3}
\end{figure}

In order to check whether the cosmic coincidence can be alleviated for this form of interaction,
we study the evolution of $r$ during the present epoch.
Since the evolution of $r$ for the case with baryons and without baryons
have nearly the same feature, we consider only the latter case.
We first consider the case where $b^2 = \lam_c = \lam_d$.
Setting $b^2$, $c$ and $\Om_{c 0}$ equal to their best fit value,
i.e. $b^2=-0.004$, $c=0.84$ and $\Om_{c 0}=0.3$,
the evolution of $r$ and $r' = \dot r / (r H)$ is plotted in figure 3.
From figure 3, we see that $|r'| > 1$ during the present epoch because
$\Om_{c 0}$ is quite different from $\Om_{c c}$.
Due to  a negative $b^2$, $r$ and $\Om_c$ become negative and consequently reach the attractor at late time.
Recall that $\Om_c > 0$ at attractor if $b^2\geq\omc / 3$.
To solve the coincidence problem, the ratio $r$ should vary slowly during the present epoch such that $|r'| \lesssim 1$ today \cite{Campo:06}.
The present value of $|r'|$ will decrease if the value of $\Om_{c c}$ gets closer to the value of $\Om_{c 0}$,
i.e. the cosmic evolution reaches the attractor near the present.
According to the yellow region in figure 1, the value of $\Om_{c c}$ will increase
and get closer to $\Om_{c 0}$ if $b^2$ increases.
However, $|r'|$ during the present epoch will be smaller than $1$ only when $b^2$ is larger than the observational bound.
The evolution of $r$ for the case where $\Om_{c c}$ is close to $\Om_{c 0}$ is shown in figure 3.
In this case, the value of $c$ and $\Om_{c 0}$ is the best fit value, while the value of $b^2$ is larger than the observational bound
but is inside the yellow region in figure 1.
From figure 3 we see that $|r'| < 1$ during the present epoch in this case
but the amount of $\Om_c$ during the early time is too small.
For this reason, this case does not satisfy the observational constraints.
Hence, in the case of $\lam_c = \lam_d$, the coincidence problem is not really alleviated for this form of interaction terms. 
\begin{figure}[ht]
\includegraphics[height=0.4\textwidth, width=0.8\textwidth,angle=0]{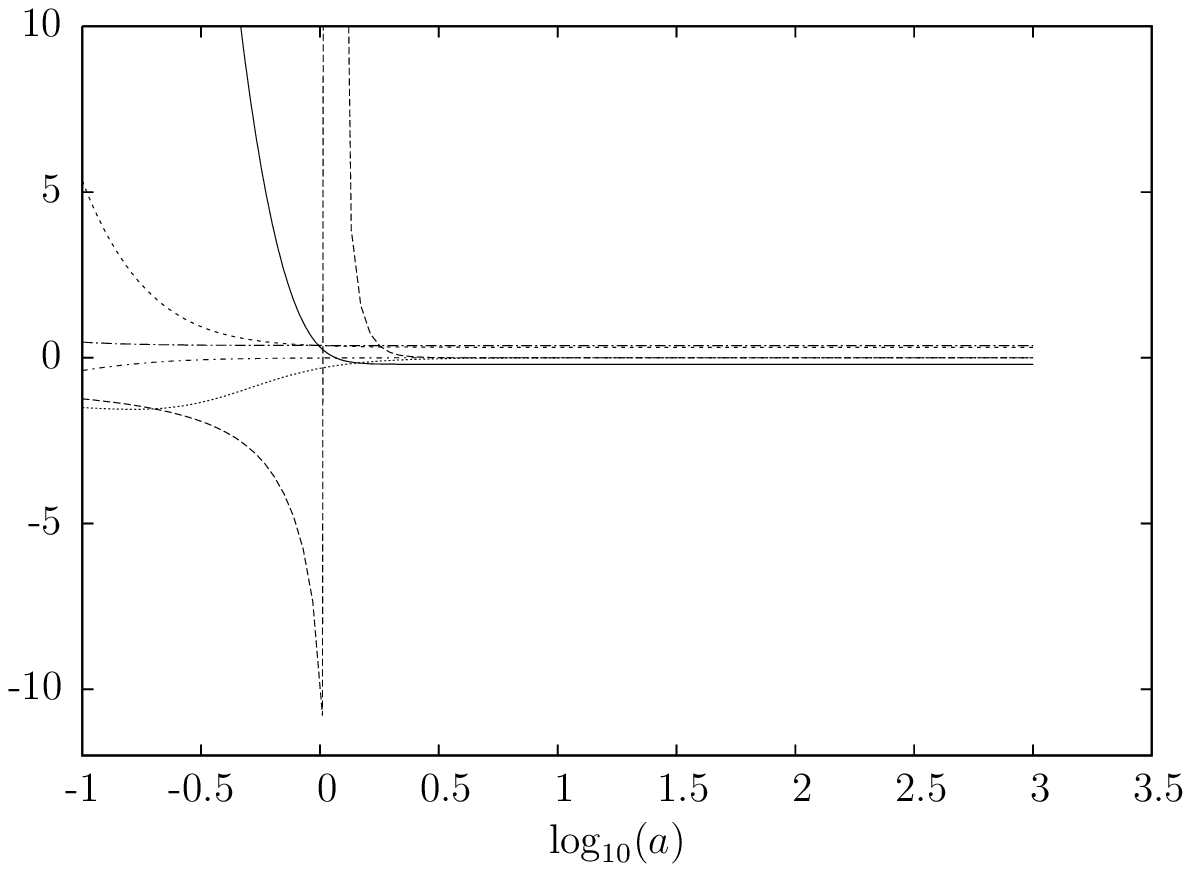}
\includegraphics[height=0.4\textwidth, width=0.8\textwidth,angle=0]{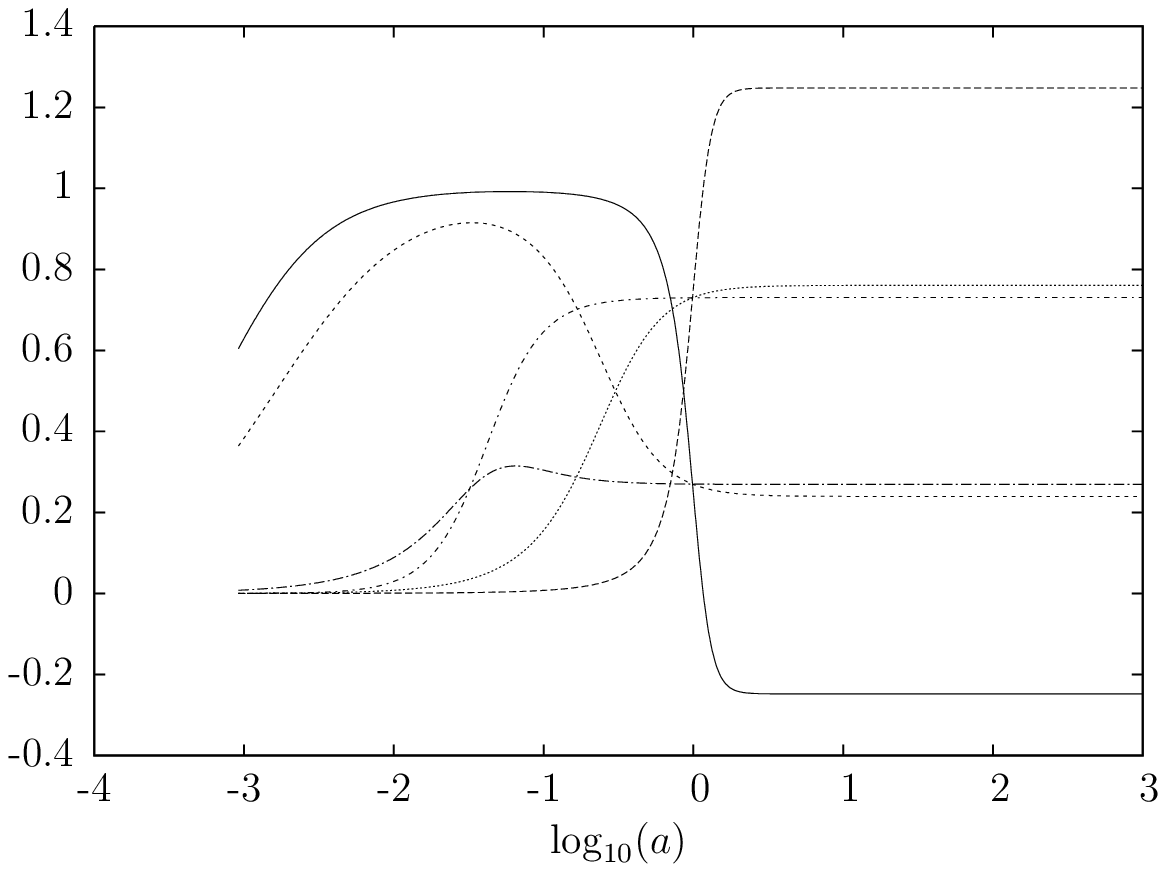}
\caption{
The upper panel shows the evolution of $r$ (solid, dashed and long dashed dotted lines) and $\dot{r}/(r H)$ (long dashed, dotted and dashed dotted lines),
while the lower panel shows the evolution of $\Om_c$ (solid, dashed and long dashed dotted lines) and $\Om_d$ (long dashed, dotted and dashed dotted lines).
The solid and long dashed lines correspond to the case where the value of $\lam_c$, $\lam_d$, $c$ and $\Om_{c 0}$ equal to their best fit value,
the dashed and dotted lines correspond to the case where $c$, $\lam_c$ and $\lam_d$ are chosen such that $\Om_{c c}$ is close to $\Om_{c 0}$,i.e. $\Om_{c c} = 0.24$,
and the long dashed dotted and dashed dotted lines correspond to the case where $c$, $\lam_c$ and $\lam_d$ are chosen such that $\Om_{c c} = \Om_{c 0}$.
}
\label{fig::4}
\end{figure}

Next, we consider the case where $\lam_c \neq \lam_d$.
We first set $c$, $\lam_c$, $\lam_d$ and $\Om_{c 0}$ equal to their best fit value,
i.e. $c=0.37$, $\lam_c = 0.08$, $\lam_d = -0.45$ and $\Om_{c 0} = 0.27$.
This choice of the parameter value is outside the yellow region in figure 2.
Since the interaction term $Q$ is dominated by $3H\lam_d\Om_d$ during the present epoch,
$Q$ becomes negative at present.
Hence, $r$ and also $\Om_c$ become negative and consequently reach the attractor at late time.
Of course, the attractor in this case does not correspond to the physical attractor.
The evolution of $r$ and $\Om_c$ is shown in figure 4.
Similarly to the case of $\lam_c=\lam_d$, the present value of $|r'|$ for this choice of parameters value is larger than
$1$ because $\Om_{c c}$ is quite different from $\Om_{c 0}$.
We now keep $\Om_{c 0}$ fixed and choose the new value of parameters $c$, $\lam_c$ and $\lam_d$ such that 
it satisfies the observational constraints and the cosmic attractor occurs at $\Om_{c c} =0.24$ near the present epoch.
This choice of the parameters value is inside the intersection between the yellow region
and the confidence region in figure 2.
From figure 4, we see that the present value of $|r'|$ is smaller than $1$ and the attractor is reached
near the present epoch for this choice of the parameters value.
Based on soft coincidence idea, the coincidence problem can be alleviated in this case.
The coincidence problem is better alleviated if $\Om_{c c} = \Om_{c 0}$
or equivalently if the cosmic attractor is reached at the present.
Unfortunately,  the parameter values that make $\Om_{c c} = \Om_{c 0}$
do not satisfy the observational constraints.
The evolution of $r$ and $\Om_c$ for this case is shown in figure 4.
From the figure, it is clear that this case is ruled out by observations.
We note that the present value of $|r'|$ will increase if we include baryons in the above consideration.

Finally, we estimate whether $|r'|$ can be smaller than $1$ at present if
we constrain the parameters of dark energy by SDSS matter power spectrum.
Here, we will not perform a complete fit for this dataset because
the nature of the perturbations in holographic dark energy is not yet completely understood.
To write down the evolution equations for the density perturbation and compute the matter power spectrum
for this interacting dark energy model, we use the assumption below.
From the results in \cite{Li:08}, we suppose that the perturbation in holographic dark energy can be neglected when $k R_e / a \gg 1$,
where $k$ is the wavenumber of the perturbation modes. Using eqs. (\ref{h}) and (\ref{holo}) and supposing that
radiation can be neglected during matter domination, one can show that $H^{-1}/R_e = \sqrt{\Om_d} /c$,
i.e. the Hubble radius is smaller than the event horizon if $\sqrt{\Om_d} < c$.
Since we use the data from SDSS which measures the matter power spectrum on scales smaller than the Hubble radius,
we neglect the perturbation in holographic dark energy in the calculation of the matter power spectrum.
We write the perturbed interaction term using the formulas in \cite{ks:84, Malik:01},
so that the evolution equations  for the perturbation in CDM are
\ba
\frac{d\Delta_c}{d\eta} &=& -k V_c + 3{\cal H}\Psi\(\lam_c + \lam_d\frac{\rho_d}{\rho_c}\)
- 3\frac{d\Phi}{d\eta}\(\lam_c + \lam_d\frac{\rho_d}{\rho_c}\) - 3{\cal H}\lam_d\Delta_c\frac{\rho_d}{\rho_c},
\nonumber \\
\frac{d V_c}{d\eta} &=& -{\cal H}V_c + k\Psi,
\ea
where $\Delta_c$ and $V_c$ are the gauge-invariant density contrast and velocity perturbation of CDM,
$\Psi$ and $\Phi$ are the metric perturbation,
${\cal H}=a^{-1}(d a/d\eta)$ and $\eta$ is the conformal time.
We solve the above equations, compute the matter power spectrum and compare the obtained matter power spectrum
with SDSS data using CMBEASY \cite{cmbeasy}.
By checking the value of $\chi^2$, we have found that the best fit parameters for this dataset are different from
the best fit parameters from the observational constraints on the homogeneous universe.
Instead of searching for the best fit parameters for this dataset,
we roughly check the viability of the parameters by comparing the matter power spectrum of the considered models
with the matter power spectrum of $\Lambda$CDM model whose parameters are taken from the  best fit value of
3-year WMAP and SDSS data \cite{wmap:06}.
In figure 5, we plot the fractional difference of the matter power spectrum between interacting holographic dark energy and $\Lambda$CDM model.
From this figure, we see that for suitable ranges of dark energy parameters, $|r'|$ can be smaller than $1$ at present
and the difference of the matter power spectrum can be smaller than the error for the matter power spectrum of SDSS data.
This implies that the alleviation of the coincidence problem by this interacting dark energy model is not excluded by SDSS data.

%
\begin{figure}[ht]
\includegraphics[height=0.4\textwidth, width=0.8\textwidth,angle=0]{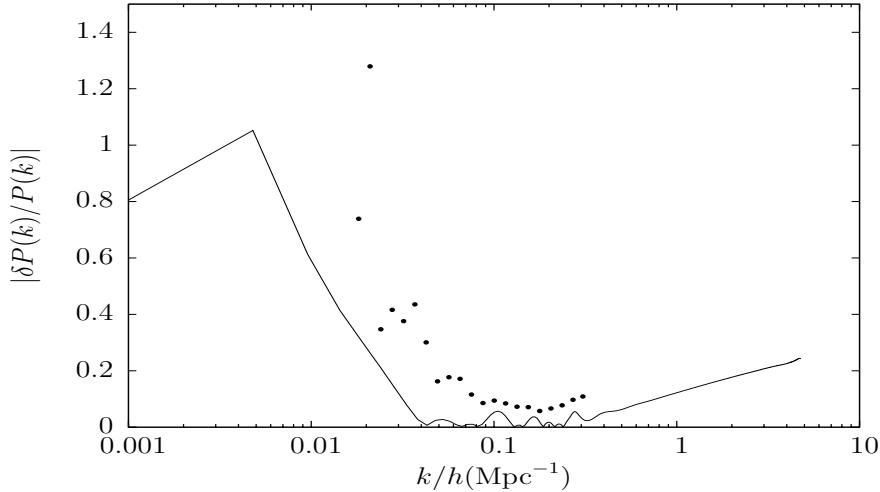}
\caption{
The fractional difference of the matter power spectrum $\left | \delta P(k)/P(k)\right |$
between interacting holographic dark energy and $\Lambda$CDM model.
For this plot, the parameters of dark energy are chosen such that $|r'|\approx 0.9$ at present.
The fractional error for the matter power spectrum $\left | \delta P(k)/P(k)\right | =$ error $/$ spectrum of SDSS data is represented by dots.
}
\label{fig::5}
\end{figure}

\section{Conclusions}

For the interacting holographic dark energy model, we study the fixed points and their stabiliby,
and compare a range of model parameters for which attractor exists with the $99.7\%$ confidence levels
from the combined analysis of SNIa data, CMB shift parameter and BAO measurement.
Neglecting baryons, the observational constraints require that
the value of $\Om_c$ at the attractor point must be small if $\lam_d = \lam_c$ or $\lam_d, \lam_c >0$. 
This implies that the cosmic evolution will reach the attractor point in the future when $\Om_c$ becomes small.
In this case, $r$ cannot be slowly varying during the present epoch and the cosmic attractor cannot be reached near the present.
Hence, the coincidence problem is not really alleviated for this case.
However, if $\lam_d$ and $\lam_c$ are allowed to be negative, the cosmic evolution can reach the attractor near
the present epoch for a narrow range of $\lam_d$ and $\lam_c$.
Therefore, the coincidence problem is  possible  to  alleviate in this case.
Including baryons in our consideration,
the attractor of the cosmic evolution cannot occur at present due to the non-vanishing baryon fraction.
According to observations, the fixed point in this case is possible only when $Q$ is small and positive.
Hence, the fixed point will be slowly reached in the future.
These results indicate that for the interacting holographic dark energy model with the interaction terms considered here, 
the cosmic coincidence problem cannot be alleviated very well.
We also briefly considered the constraint from SDSS matter power spectrum on the dark energy parameters.
We have found that the parameters ranges that lead to the alleviation of cosmic coincidence are allowed by SDSS data.

\section*{Acknowledgments}

The author would like to thank Bin Wang for helpful discussions and suggestions and for
hospitality at Fudan University during the final stage of this work.
He also thanks A. Ungkitchanukit and anonymous referee for comments on the manuscript.
The observational fitting was performed using the workstation
of the Biophysics group, Kasetsart University.
This work is supported by Thailand Research Fund (TRF) and vcharkarn.com.

\section*{References}

\providecommand{\href}[2]{#2}\begingroup\raggedright
 
\end{document}